\documentclass[fleqn,twoside]{niaauthm}

\usepackage{graphicx}

\begin{document}

\begin{frontmatter}

\title{High sensitivity measurement of $^{224}$Ra and $^{226}$Ra in water \\ with an
improved hydrous titanium oxide technique \\ at the Sudbury Neutrino Observatory}

\author[Laurentian]{B.~Aharmim},
\author[{Oxford,Queens,SNOLAB}]{B.T.~Cleveland},
\author[{Carleton,Oxford,Queens}]{X.~Dai}$^{,1}$,
\thanks{Corresponding author. Tel: 1-613-533-2694; 
        Fax: 1-613-533-6813; E-mail: xdai@owl.phy.queensu.ca.}
\author[Oxford]{G.~Doucas},
\author[Laurentian]{J.~Farine},
\author[Oxford]{H.~Fergani},
\author[SNOLAB]{R.~Ford},
\author[Brookhaven]{R.L.~Hahn},
\author[Laurentian]{E.D.~Hallman},
\author[Oxford]{N.A.~Jelley},
\author[Brookhaven]{R.~Lange},
\author[Oxford]{S.~Majerus},
\author[Carleton]{C.~Mifflin},
\author[Queens]{A.J.~Noble},
\author[Oxford]{H.M.~O'Keeffe},
\author[Laurentian]{R.~Rodriguez-Jimenez},
\author[Carleton]{D.~Sinclair},
\and
\author[Brookhaven]{M.~Yeh}

\address[Brookhaven]{Chemistry Department, Brookhaven National Laboratory,
         Upton, New York 11973-5000, USA}
\address[Carleton]{Ottawa-Carleton Institute for Physics, Department of Physics, 
         Carleton University, Ottawa, Ontario K1S 5B6, Canada}
\address[Laurentian]{Department of Physics, Laurentian
         University, Sudbury, Ontario P3E 2C6, Canada}
\address[Oxford]{Department of Physics, University of Oxford, Denys Wilkinson Building,
         Keble Road, Oxford, OX1 3RH, UK}
\address[Queens]{Department of Physics, Queen's University, Kingston,
         Ontario K7L 3N6, Canada}
\address[SNOLAB]{SNOLAB, Sudbury, Ontario P3Y 1M3, Canada}

\begin{abstract}

The existing hydrous titanium oxide (HTiO) technique for the measurement of $^{224}$Ra and $^{226}$Ra in the water at the
Sudbury Neutrino Observatory (SNO) has been changed to make it faster and less sensitive to trace impurities in the HTiO
eluate. Using HTiO-loaded filters followed by cation exchange adsorption and HTiO co-precipitation, Ra isotopes from 200-450
tonnes of heavy water can be extracted and concentrated into a single sample of a few millilitres with a total chemical
efficiency of 50\%. Combined with beta-alpha coincidence counting, this method is capable of measuring 1.5$\times$10$^{-3}$
$\mu$Bq/kg of $^{224}$Ra and 3.3$\times$10$^{-3}$ $\mu$Bq/kg of $^{226}$Ra from the $^{232}$Th and $^{238}$U decay chains,
respectively, for a 275 tonne D$_{2}$O assay, which are equivalent to 4$\times$10$^{-16}$ g Th/g and
3$\times$10$^{-16}$ g U/g in heavy water. 

\end{abstract}


\maketitle
\end{frontmatter}


\section{Introduction}

SNO is a heavy water Cherenkov detector situated at a depth of about 2100 m in
Vale Inco's Creighton
mine in Sudbury, Canada \cite{SNO1}. It was built to detect neutrinos from the Sun and to understand
the origin of the ``Solar Neutrino Problem'' - the longstanding disagreement between the measured
neutrino flux from several previous solar neutrino experiments \cite{SNE1,SNE2,SNE3,SNE4,SNE5,SNE6}
and theoretical predictions from the Standard Solar Models (SSM) \cite{SSM1}. The SNO detector
consists of an inner neutrino target of 1000 tonnes of ultrapure D$_{2}$O held in a 12 m diameter
acrylic sphere which is surrounded by a shield of 7000 tonnes of H$_{2}$O contained in a 34 m high
barrel-shaped cavity of maximum diameter 22 m. An array of approximately 9500 inward-looking
photomultiplier tubes, mounted on a geodesic sphere of diameter 17.8 m, detects the Cherenkov light produced as a result of neutrino interactions occurring in the D$_{2}$O. 

Solar neutrinos are detected in SNO through three distinct reactions: the charged current (CC)
reaction, $\nu$$_{e}$ + d $\rightarrow$ p + p + e$^{-}$; the neutral current (NC) reaction,
$\nu$$_{x}$ + d $\rightarrow$ p + n + $\nu$$_{x}$, ({$x$ = e, $\mu$, $\tau$); and the elastic-scattering (ES) reaction,
$\nu$$_{x}$ + e$^{-}$ $\rightarrow$ $\nu$$_{x}$ + e$^{-}$.  The CC reaction is only sensitive to electron
neutrinos --- the neutrino type produced by fusion reactions in the Sun, whereas the NC reaction is equally sensitive to all active neutrino flavours. 

There were three phases in the SNO experiment: pure D$_{2}$O (November 1999 to April 2001), salt (May
2001 to September 2003) when 2 tonnes of NaCl were added to the heavy
water to improve the detection of neutrons from  neutrino NC interactions, and the Neutral Current Detector (NCD) phase (November 2004 to November 2006). 
The total flux of all active neutrinos measured during the first two phases of the experiment agreed very well with the predictions made by the SSM,
while only one third of the expected solar neutrinos were detected by the CC reaction
\cite{SNO2,SNO3,SNO4,SNO5}.  This provided direct evidence that neutrino flavour change is the cause of the Solar Neutrino Problem. 

The water used in the SNO detector had to be extremely clean to minimize background signals from naturally occurring radioactivity. The
$^{232}$Th and $^{238}$U decay chains both contain gamma rays with energies greater than 2.22 MeV. These gammas produced by $^{208}$Tl
from the $^{232}$Th chain and $^{214}$Bi from the $^{238}$U chain could photodisintegrate a deuterium nucleus, producing a
neutron that was
indistinguishable from a NC event.  Radiopurity limits in the water were set by constraining the backgrounds from each decay chain to be less than one
photodisintegration neutron per day in the SNO detector, compared with $\sim$12.5 neutrons per day from solar
neutrino NC interactions. 

To measure the amounts of $^{208}$Tl and $^{214}$Bi, SNO developed three separate techniques to assay and purify $^{224}$Ra (the
$^{208}$Tl precursor), and $^{226}$Ra and $^{222}$Rn (the $^{214}$Bi precursors). These were MnOx \cite{MnOx} and HTiO \cite{HTiO} for
$^{224}$Ra and $^{226}$Ra, and degassing for $^{222}$Rn \cite{Rn}. The $^{232}$Th and $^{238}$U chains were not in secular equilibrium
in SNO water due to leaching of radium isotopes from trace contamination of materials in contact with the water and from
ingress of $^{222}$Rn (half-life of 3.8 d). 
However, to facilitate comparisons between different assay methods, concentrations are quoted in terms of the equivalent amounts
of $^{232}$Th and $^{238}$U in a chain in equilibrium. For a limit of one photodisintegration neutron per day the equivalent 
limits on $^{232}$Th and $^{238}$U were 3.8$\times$10$^{-15}$ g Th/g and 3.0$\times$10$^{-14}$ g U/g
in the heavy water. The requirement in the light water surrounding the heavy water was less stringent, with upper limits of
3.7$\times$10$^{-14}$ g Th/g water and of 4.5$\times$10$^{-13}$ g U/g water.

In the original HTiO D$_{2}$O assay procedure \cite{HTiO}, HTiO adsorbent was deposited onto a pair of 1 m long filters. The loaded filters were connected to the detector water systems underground and Ra from a few hundred tonnes of D$_{2}$O was
extracted.  Once the assay was complete, the Ra was removed by eluting the filters with 15 L of 0.03 mol/L of
HNO$_{3}$. In the subsequent secondary concentration step, the HNO$_{3}$ eluate was further concentrated to a
small sample volume for counting by co-precipitating radium with HTiO and passing it through a series 
of three MediaKap-10 filters; each filter was then eluted three times to ensure all of the activity
was removed. The three eluates from each MediaKap-10 filter were combined, thus producing three samples
per assay.  This method was time consuming and required many samples to be measured in the beta-alpha
counting system to determine the amount of radium extracted.  

During the salt phase of the SNO experiment, a significant decrease in the secondary concentration efficiency was observed. 
This was attributed to increased levels of Mn in the D$_{2}$O from leaching of the beads used in the MnOx
technique, which was extracted by the HTiO columns, forming 
hydrous manganese oxide during the co-precipitation step in the secondary concentration stage.
The hydrous manganese oxide absorbed some of the Ra present in the sample, which could not be removed during the subsequent HCl elution of the MediaKap filters, resulting in a decreased secondary concentration efficiency. 

In addition, Ni was expected to leach slightly from the NCD detectors that were deployed in the D$_{2}$O
region. Levels of less than about 1 ppb Ni were expected (and found) in the D$_2$O during the NCD phase.
While this concentration would not affect the optical properties of
the SNO detector it could, similar to manganese, decrease the secondary concentration efficiency of the HTiO
method. Therefore, the effect of Ni on the new technique
was studied.
 
To overcome the problem of interference by trace elements in the HTiO eluate, new elution and secondary
concentration procedures
were developed. This method is faster than the original method and it also reduces the number of samples produced and
counted for each assay. In this paper, the details of the new method, which was used in the NCD phase of SNO, and evaluation of its efficiency are presented. All modifications are 
equally applicable to both light and heavy water assays.

\section{The new HTiO procedure}

An overview of the HTiO assay procedure is shown in Fig.~1. It consists of five steps: HTiO deposition, Ra extraction,
elution, secondary concentration and counting. The HTiO adsorbent, formed as a white colloidal suspension by hydrolysing
Ti(SO$_{4}$)$_{2}$, is deposited onto a Memtrex pleated filter (Osmonics, Inc., USA) made from polyethersulfone or polypropylene (PP) with a Ti coverage of 2.5 g/m$^{2}$.
To achieve this, 15 L of dilute HTiO solution is recirculated through the filter
at approximately 80 L/min for 10 min. A typical D$_{2}$O assay used a pair of 1 m long filters each with a surface area of 2.08 m$^{2}$ and a pore size of 0.1 $\mu$m. For H$_{2}$O assays, 0.25 m long filters were used with a surface area of 0.52 m$^{2}$ and a pore size of 0.1 $\mu$m. The filters used in D$_{2}$O assays must be deuterated before use in an assay to maintain the isotopic purity of the SNO heavy water. The HTiO production, deposition and deuteration procedures are described in detail in \cite{HTiO}.

The filters are housed in PP columns and are loaded with HTiO before being transported underground.  For an assay, 
the filters are attached to the SNO water circulation system where between 200 and 450 tonnes of D$_{2}$O
(20-40 tonnes of H$_{2}$O) are passed through them at a typical rate of 22 L/(min$\cdot$m$^{2}$).
After extraction, Ra is eluted from the HTiO-loaded filters by circulating 15 L of 0.1 mol/L HCl through the columns 
at a flow rate of approximately 80 L/min for 20 min. The use of 0.1 mol/L HCl, rather than 0.03 mol/L
HNO$_{3}$ as in the original procedure, is because the cation exchange resin used in the subsequent step extracts Ra in HCl
with a higher efficiency; it also removes Th as well as Ra from the HTiO-loaded filters.
In the secondary concentration step, Ra (and any Th if present) is extracted onto a cation exchange resin by
flowing the 15 L HTiO eluate at a rate of 250 mL/min through a polymethylpentene (PMP) column (1.0 cm diameter, 18.0 cm length) containing
12.0 g of Dowex 50WX8 resin (H$^{+}$-form, 100 mesh). The resin is mixed with 50 mL of 0.1 mol/L
HCl and added to the column, followed by rinsing with 50 mL of 0.1 mol/L HCl before the eluate is passed through it.
To minimise the procedural background, the resin is only used once for each assay or background experiment.

The resin column is rinsed with ultrapure water (UPW) to neutral pH and then turned in reverse flow
direction for Ra elution. Approximately 100 mL
of 0.25 mol/L of disodium ethylenediaminetetraacetate (EDTA) is passed through the column
at pH 10 and a flow rate of 5 mL/min to remove Ra. About 15\% of any Th extracted by the resin is also removed. The
EDTA eluate is collected in a PTFE beaker. Upon addition of 8 mL of concentrated HNO$_{3}$ to the eluate,
the EDTA precipitates. This EDTA is decomposed by boiling the solution for 15 minutes after which time a
second 8 mL of concentrated HNO$_{3}$ is added, and boiling is continued for a further 15 minutes. The next step involves co-precipitation with HTiO. Approximately 
250 mL of UPW is added to the residual solution along with 2 mL of 15\% Ti(SO$_{4}$)$_{2}$ (60 mg of Ti). This solution is
titrated to pH 9 with NaOH which causes co-precipitation of the HTiO and Ra. In the final step, the solution is centrifuged at
3400 rpm for 3 min. The supernatant is decanted and the HTiO precipitate is dissolved in 1.5 mL of concentrated HCl. This
final sample is diluted with UPW to 8 mL and mixed with 42 g of Optiphase HiSafe 3 liquid scintillator cocktail (purchased
from PerkinElmer, Inc.) in a 60 mL PMP jar. The amount of $^{224}$Ra and $^{226}$Ra is inferred by counting the samples using a beta-alpha coincidence counter for 10-14 days. 

The beta-alpha coincidence counting system is described in detail in \cite{HTiO}.  The background of this
counting system was measured using a blank sample (a mixture of 10 mL of 0.5 mol/L HCl and 42 g of liquid scintillator cocktail) and the count rate was found to be very low, 0.03 h$^{-1}$ for $^{224}$Ra and 0.3 h$^{-1}$ for $^{226}$Ra. The $^{226}$Ra background is mainly contributed by the Optiphase HiSafe 3 liquid scintillator cocktail. The counting efficiencies were also measured to be 45$\pm$5\% for $^{224}$Ra and 60$\pm$10\% for $^{226}$Ra. These efficiencies were regularly checked using $^{226}$Ra and $^{228}$Th calibration sources.
Throughout the paper $\pm$ values are standard combined uncertainties (k=1).

\begin{figure}[ht]
\begin{center}
\includegraphics[width=70mm]{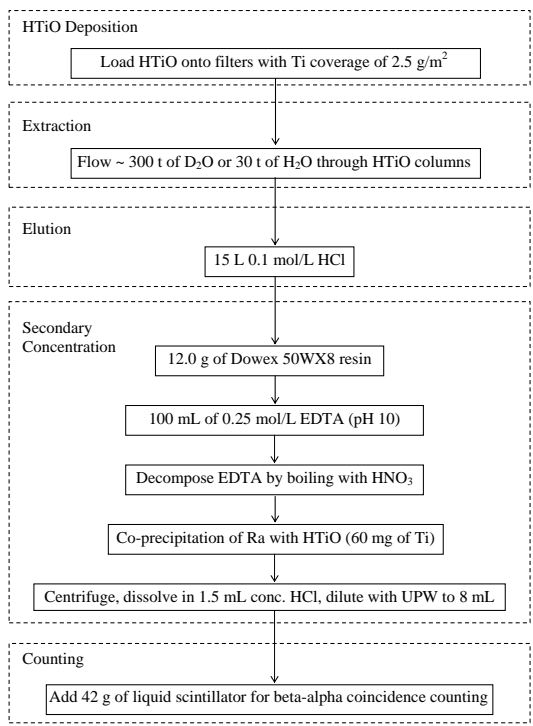}
\caption{The new Ra procedure for an HTiO water assay.}
\label{fig:Procedure}
\end{center}
\end{figure}

\section{Results and discussion}

\subsection{Efficiencies for the new procedure}

The Ra efficiencies for the new procedure were determined by spike experiments using 0.01-0.5 Bq of $^{226}$Ra tracer and are given,
along with the uncertainties, in Table 1. 
The total chemical recovery for Ra was determined to be 50$\pm$8\%. Combined with the counting efficiencies, the total procedural efficiencies are 30$\pm$7\% for $^{226}$Ra and 22$\pm$4\% for $^{224}$Ra. In the following sections, the methods used to obtain the efficiency for each step are discussed in detail.
 
\begin{table}
\caption{Summary of efficiencies of Ra for the new procedure. }
\label{table:1}
\newcommand{\m}{\hphantom{$-$}}
\newcommand{\cc}[1]{\multicolumn{1}{c}{#1}}
\renewcommand{\tabcolsep}{9pt} 
\renewcommand{\arraystretch}{1.2} 
\begin{tabular}{@{}lll} \hline 
Efficiency & $^{226}$Ra & $^{224}$Ra \\ \hline 
Extraction $\varepsilon$$_{ext}$ & \multicolumn{2}{c}{95$\pm$5\%} \\ 
Elution $\varepsilon$$_{elu}$ & \multicolumn{2}{c}{90$\pm$10\%} \\  
Secondary concentration $\varepsilon$$_{conc}$ & \multicolumn{2}{c}{58$\pm$6\%} \\ 
Total chemical ($\varepsilon$$_{ext}$$\cdot$$\varepsilon$$_{elu}$$\cdot$$\varepsilon$$_{conc}$) &
\multicolumn{2}{c}{50$\pm$8\%} \\ 
Counting  $\varepsilon$$_{count}$ & 60$\pm$10\% & 45$\pm$5\% \\ 
Total ($\varepsilon$$_{ext}$$\cdot$$\varepsilon$$_{elu}$$\cdot$$\varepsilon$$_{conc}$$\cdot$
$\varepsilon$$_{count}$) & 30$\pm$7\% & 22$\pm$4\% \\ \hline 
\end{tabular}\\[2pt]
\end{table}

\subsubsection{Extraction}

In previous studies \cite{HTiO,Heron}, the extraction efficiency for Ra was 
carefully examined, and for a typical HTiO water assay in the pure D$_{2}$O phase ($\sim$200 tonnes for D$_{2}$O), the extraction efficiency was determined to be 95$\pm$5\% at a flow rate of 19 L/(min$\cdot$m$^{2}$). This was measured by flowing D$_{2}$O through two sets of HTiO-loaded filters (each set consisting of two parallel filters), connected in series.  The activities extracted by the two sets were compared. 
The formula used to calculate the extraction efficiency, $\varepsilon$$_{ext}$, is 
\\
$\varepsilon$$_{ext}$ = 1 -- $\frac{A_{d}}{A_{u}}$ ,
\\
where $A_{d}$ and $A_{u}$ are the activities on the downstream and upstream columns,
respectively, and where the first two columns are the upstream columns. 

During the NCD phase the heavy water was passed through a reverse osmosis unit from time to time 
to maintain a low level of metallic ions as in the pure D$_{2}$O phase, so an extraction of 95$\pm$5\% was assumed. Checks of the extraction efficiency made during the NCD phase were consistent with this value.

\subsubsection{Elution}

To investigate the effect of HCl concentration on the Ra and Th elution efficiencies, several small scale spike
experiments were carried out using HTiO-loaded (2.5 g Ti/m$^{2}$) MediaKap filters. These have identical pore size, but
smaller surface area (0.015 m$^{2}$) than the Memtrex filters used for assays. The elution efficiency of Ra was found
to be 90$\pm$10\% for HCl concentrations ranging from 0.1-0.5 mol/L, while the elution
efficiency of Th was decreased from 95$\pm$5\% to 65$\pm$5\% with HCl concentrations decreasing from 0.5 to 0.1 mol/L. Higher concentration 
of HCl would also dissolve more Ti from the loaded filters which could reduce the Ra extraction efficiency by the
cation exchange resin in the secondary concentration step. Therefore, 0.1 mol/L of HCl was chosen to elute Ra and Th
from the HTiO filters. The Ra elution efficiency for this acidity
was determined to be 90$\pm$10\% which is a slight improvement on the use of HNO$_{3}$ in the previous
method. 

\subsubsection{Secondary concentration}

One major concern for the secondary concentration efficiency of water assays was
the presence of impurities such as Mn and Ni in the water, which could be observed in the HTiO eluate along with Ti stripped from the HTiO-loaded filters 
during elution. By adding a known amount of $^{226}$Ra and $^{228}$Th radiotracers, spike experiments
were conducted to determine the efficiency for each step of the new secondary concentration procedure
in the presence of Mn, Ni and Ti. In each spike experiment, 130 mg of Mn, 140 mg of Ni and 1500 mg of
Ti were added to 15 L of 0.1 mol/L HCl to mimic a typical assay eluate. The amounts of Mn and Ni were
based on the estimated maximum adsorption capacities for these elements by a pair of 1 m long HTiO
loaded filters measured in an assay during the transition from the salt to the NCD phase of SNO. The amount of Ti was based upon test elutions of HTiO-loaded filters. In these spike experiments, 
$^{226}$Ra and $^{228}$Th were counted using the beta-alpha coincidence counters, while all of the stable
tracers, including Ba, Mn, Ni and Ti, were measured using inductively coupled plasma mass spectrometry (ICP-MS) 
at the Geoscience Laboratories, Sudbury, Ontario.

The secondary concentration procedure consists of two steps. In the first step, Ra is adsorbed onto an ion exchange (IX) resin and
eluted into a relatively small volume. The second step further concentrates Ra, resulting in an 8 mL sample. The Dowex 50WX8 cation
exchange resin is widely used for the pre-concentration of Ra in environmental samples \cite{ALHA,DITCH,HIGU} due to its strong
retention ability and fast adsorption equilibrium. The adsorption and elution behaviour of Ra from this resin were carefully examined
for extraction of Ra from 15 L of the HCl eluate.  The extraction efficiency for Ra by the IX resin was measured to be greater than
90\% in the presence of Mn, Ni and Ti from the spike experiments (see Table 2) (the Ra is preferentially
adsorbed by both the HTiO-loaded filter during extraction and by the IX resin over Mn, Ni or Ti). The
elution of Ra from the IX resin at various EDTA concentrations and pHs was also studied. It was found that
90\% of the extracted Ra was eluted into 100 mL of 0.25 mol/L EDTA at pH 10. Most of the Ti
remained on the resin (see Table 2). An HTiO co-precipitation method is used to further concentrate the Ra
from 100 mL of EDTA eluate for beta-alpha coincidence counting. Since co-precipitation would not occur in the presence of a high concentration of EDTA, the EDTA must be decomposed by boiling with concentrated HNO$_{3}$ prior to this step. The spike experiments indicated that an average of 69\% of the eluted Ra from the resin could be concentrated into the final counting sample (see Table 2). Tests were also performed in the absence of Ni and Mn and no observable difference in the secondary concentration efficiency was found.

\begin{table*}[htb]
\caption{Efficiencies of Ra, Th, Ni, Mn and Ti in the secondary concentration step from the spike
experiments.}
\label{table:2}
\newcommand{\m}{\hphantom{$-$}}
\newcommand{\cc}[1]{\multicolumn{1}{c}{#1}}
\renewcommand{\tabcolsep}{16pt} 
\renewcommand{\arraystretch}{1.2} 
\begin{tabular}{@{}lccccc} \hline 
 & \multicolumn{3}{c}{Efficiencies in secondary concentration step } & \multicolumn{2}{c}{Impurity contents (mg)} \\  
 & IX extraction & IX elution & HTiO co-precipitation & HCl eluate & Final sample \\ \hline 
$^{226}$Ra & 93$\pm$4\% & 90$\pm$6\% & 69$\pm$5\% &  &  \\ 
$^{228}$Th & 98$\pm$2\% & 15$\pm$2\% & 48$\pm$10\% &  &  \\ 
Ba & 94$\pm$4\% & 95$\pm$5\% & 15$\pm$1\% &  &  \\ 
Ni & 33$\pm$3\% & 67$\pm$7\% & $<$0.2\% & 140 & $<$0.06 \\ 
Mn & 33$\pm$3\% & 75$\pm$9\% & $<$7\% & 130 & $<$3 \\ 
Ti & 51$\pm$9\% & 3.0$\pm$0.4\% & $\sim$75\% & 1500 & $\sim$62 \\ \hline 
\end{tabular}\\[2pt]
\end{table*}

\subsubsection{Counting}

In principle high amounts of Mn or Ni could reduce the counting efficiency as both have absorption bands between 350 and 500 nm, the region of wavelengths 
to which the photomultiplier tubes used in the beta-alpha coincidence counters are most sensitive.  However, as only very small quantities of Mn and Ni 
were present in the final sample, the counting efficiency was unaffected. 
A typical assay sample looks milky white after mixing with liquid scintillator cocktail due to up to $\sim$62 mg Ti in the final sample. This does not affect the counting efficiency significantly as the scintillation light is only scattered but not absorbed; as a result the position of the alpha peak is not altered appreciably. Tests were performed to verify this and no obvious decrease in counting efficiency was observed (see Fig.~2). The counting efficiencies were found to be $45\pm5$\% for $^{224}$Ra and $60\pm10$\% for $^{226}$Ra.  Combining all of the individual efficiencies, the total secondary concentration efficiency 
is found to be 58$\pm$6\% for Ra (see Table 1).

\begin{figure}[ht]
\begin{center}
\includegraphics[width=70mm]{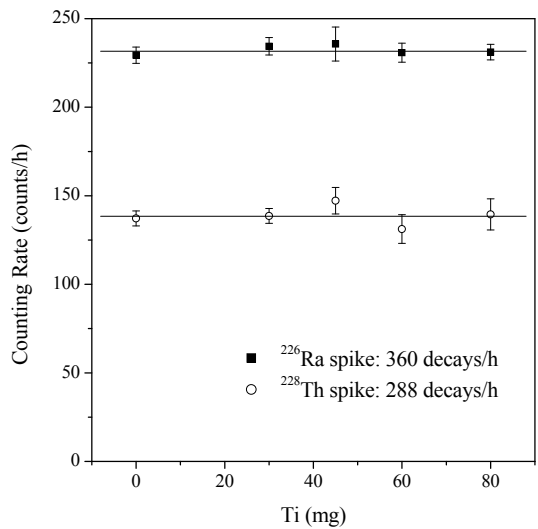}
\caption{Effect of various Ti concentrations on the counting efficiencies.(Sample matrix: titanium, 
in the form of HTiO was dissolved in 2.0 mL of concentrated HCl + 6.0 mL of UPW + 42 g of 
Optiphase HiSafe 3 liquid scintillation cocktail.)}
\label{fig:Quench}
\end{center}
\end{figure}

\subsubsection{$^{228}$Th measurement}

As the HCl elution of the HTiO-loaded filters removes Th with an efficiency of 65$\pm$5\%, Th will
be extracted by the IX resin, along with the Ra. Approximately 15\% of the Th is removed from the IX
resin when the Ra is eluted with EDTA (see the IX elution efficiency for $^{228}$Th in Table 2). Following the EDTA elution the column is sealed and left moist
and Ra builds up from the Th that remains on the resin. Approximately 9-10 days after the initial
EDTA elution, the equilibrium between $^{228}$Th and $^{224}$Ra is well established and the resin may
be re-eluted with EDTA.  The secondary concentration proceeds as before, and the efficiencies of all
subsequent stages are identical to those given in Tables 1 and 2. As the elution efficiency is lower 
for Th compared to Ra and as approximately 15\% of the Th is removed with the initial IX elution, 
the overall chemical efficiency is approximately 27\% for a second IX elution.

\subsection{Background and detection limit}

To determine the contribution of the equipment and reagents to the radioactive signal observed for an assay, a background
measurement must always be carried out. This takes place typically one week before the assay. For a background measurement,
the filters are loaded with HTiO and processed in an identical manner to an assay. Underground extraction does not occur
during the background measurement. Following this measurement the filters are cleaned using 0.5 mol/L HCl and re-loaded with
HTiO for the assay. The count rate for a typical background signal was found to be 0.25$\pm$0.1 h$^{-1}$ for $^{224}$Ra ($^{232}$Th chain) and 0.85$\pm$0.3 h$^{-1}$ for $^{226}$Ra ($^{238}$U chain). 
This background arises from Ra leaching from the apparatus and from the reagents used in the elution and secondary
concentration processes. Taking the detection limit as 3.29 times the standard deviation of the background rate \cite{Currie},
the minimum detectable concentrations of the new
procedure for a 275 tonne assay are 1.5$\times$10$^{-3}$
$\mu$Bq/kg of $^{224}$Ra and 3.3$\times$10$^{-3}$ $\mu$Bq/kg of $^{226}$Ra, which are equivalent to $\sim$4$\times$10$^{-16}$ g Th/g and
$\sim$3$\times$10$^{-16}$ g U/g in heavy water.

\subsection{Assay results}

A 275 tonne assay of the heavy water in the NCD phase of SNO gave a (signal minus background) count rate of 0.5$\pm$0.2
h$^{-1}$ for $^{224}$Ra ($^{232}$Th chain) and 0.0$\pm$0.4 h$^{-1}$ for $^{226}$Ra ($^{238}$U chain). 
These were the rates at the start of counting. The Ra activity in the water was deduced after allowing 
for the time of extraction ($\sim$3 d) and the delay between the end of extraction and counting ($\sim$5-6 h). Assuming radioactive
equilibrium in the $^{232}$Th and $^{238}$U chains, these rates correspond to a value of
8$\pm$4$\times$10$^{-16}$ g Th/g water and less than 3$\times$10$^{-16}$ g U/g water. 
The systematic error arises mainly through the uncertainties of $\sim$20\% for $^{224}$Ra and $\sim$24\% for $^{226}$Ra
in the efficiencies of the chemical procedures and counting (see Table \ref{table:1}), and to a lesser extent from the
uncertainties in the background subtraction and in the volume of water sampled. The systematic uncertainty
contributions (k=1) are 
estimated to be $\pm$24\% for the $^{232}$Th chain and $\pm$29\% for the $^{238}$U chain. This procedure was used for light water as well as heavy water HTiO assays in the NCD phase of the SNO experiment.

The new procedure was also used to assay trace amounts of $^{224}$Ra and $^{226}$Ra on the surface of sections of one of the NCDs.  Each section was placed in a column, connected to the elution rig and the outer surface rinsed with 0.1 mol/L HCl at 80 L/min for 20 minutes.  Following the elution, the concentration
of Ni in the 15 L eluate was very high (approximately 80 mg of Ni was present). The new secondary concentration procedure was followed to 
extract Ra.  To ensure that the secondary concentration efficiency was unchanged, the procedure was tested using a number of
test sections which had been plated with $^{228}$Th. The secondary concentration efficiency derived
from these tests was in agreement with that given in Table \ref{table:1}.  The final results were 
0.91$\pm$0.21 $\mu$g of $^{232}$Th and 0.11$\pm$0.03 $\mu$g of $^{238}$U,
in good agreement with measurements obtained by two other independent techniques \cite{NCD}.

\section{Conclusions}

We have developed and implemented a new secondary concentration stage for the HTiO-based assay procedure used to determine the Ra concentration in both the light and the heavy water of the SNO detector. Compared to the previous HTiO assay method \cite{HTiO}, the new procedure offers the following advantages:

\noindent
(1) It is less sensitive to impurities in the HTiO eluate and was unaffected by concentrations of
$9.3\times 10^{-6}$ g Ni/mL, $8.7
\times 10^{-6}$ g Mn/mL and $1.0\times 10^{-4}$ g Ti/mL in the 15 L HTiO eluate.

\noindent
(2) It requires fewer beta-alpha coincidence counters; only two samples (one for assay plus one for background) are produced for one water assay using the new procedure while six samples (three assay, three background) were produced using the previous method.

\noindent
(3) The time required to process the filters once the assay has been completed 
 is reduced from 12 hours to around 5-6 hours.

The new HTiO procedure has been used to assay the light and heavy water of the SNO detector, with detection limits of 1.5$\times$10$^{-3}$
$\mu$Bq/kg for $^{224}$Ra and 3.3$\times$10$^{-3}$ $\mu$Bq/kg for $^{226}$Ra, which are equivalent to 4$\times$10$^{-16}$ g Th/g and
3$\times$10$^{-16}$ g U/g for a 275 tonne heavy water 
assay. It has also been used to assay for trace amounts of
$^{224}$Ra and $^{226}$Ra on sections of the NCDs. With appropriate validations, the technique could have a variety of applications in the measurement
of low-level natural radioactivity in different environmental and biological samples.

\section{Acknowledgements}

The authors would like to thank Carol Woodliffe for making the HTiO, Mike Tacon and the Oxford
mechanical workshop, Wing Lau and the Oxford design office and the SNO site U/G operations crew for
all their contributions. We would also like to thank Vale Inco Ltd and their staff at the Creighton mine for their cooperation and Atomic Energy of Canada Ltd (AECL) for the generous loan of the heavy water in cooperation with Ontario Power Generation.

This work was supported in the United Kingdom by the Science and Technology
Facilities Council (formerly the Particle Physics and Astronomy Research
Council); in Canada by the Natural Sciences and Engineering Research Council,
the National Research Council, Industry Canada, the Northern Ontario Heritage
Fund Corporation, and the Province of Ontario; in the USA by the Department of
Energy. Further support was provided by Vale Inco Ltd., AECL, Agra-Monenco, Canatom, the Canadian Microelectronics Corporation, AT\&T Microelectronics, Northern Telecom, and British Nuclear Fuels, Ltd.

\end{document}